# Few-layer $MoS_2$ flakes as hole-selective layer for solution-processed hybrid organic hydrogen-evolving photocathodes

Sebastiano Bellani,[a][†] Leyla Najafi,[a,b][†] Andrea Capasso,[a] Antonio Esau Del Rio Castillo,[a] Maria Rosa Antognazza,[c] and Francesco Bonaccorso[a][*]

[a] Graphene Labs, Istituto Italiano di Tecnologia, via Morego 30, 16163 Genova, Italy.

[b] Università degli studi di Genova, Dipartimento di Chimica e Chimica Industriale, Via Dodecaneso 31, 16163 Genova, Italy.

[c] Center for Nano Science and Technology @Polimi, Istituto Italiano di Tecnologia, Via Pascoli 70/3, 20133 Milano, Italy.

† These authors contributed equally.

* Corresponding authors: francesco.bonaccorso@iit.it.

## Abstract

High-efficiency organic photocathodes, based on regioregular poly(3-hexylthiophene) and phenyl-C61-butyric acid methyl ester (rr-P3HT:PCBM) bulk heterojunction sandwiched between charge-selective layers, are emerging as efficient and low-cost devices for solar hydrogen production by water splitting. Nevertheless, stability issues of the materials used as charge-selective layers are hampering the realization of long-lasting photoelectrodes, pointing out the need to investigate novel and stable materials. Here, we propose $MoS_2$ nano-flakes, produced by Li-aided exfoliation of bulk counterpart, as efficient atomic-thick hole-selective layer for rr-P3HT:PCBM-based photocathodes. We carried out a p-type chemical doping to tune on-demand the $MoS_2$ Fermi level in order to match the highest occupied molecular orbital level of the rr-P3HT, thus easing the hole collection at the electrode. The as-prepared p-doped $MoS_2$-based photocathodes reached a photocurrent of 1.21 mA $cm^{-2}$ at 0 V *vs.* RHE, a positive onset potential of 0.56 V *vs.* RHE and a power-saved figure of merit of 0.43%, showing a 6.1-fold increase with respect to pristine $MoS_2$-based photocathodes, under simulated 1 Sun illumination. Operational activity of the photocathodes over time and under 1 Sun illumination revealed a progressive stabilization of the photocurrents at 0.49 mA $cm^{-2}$ at 0 V *vs.* RHE. These results pave the way towards the exploitation of layered crystals as efficiency-boosters for scalable hybrid organic $H_2$-evolving photoelectrochemical cells.

## 1. Introduction

Sunlight and water are the most abundant, clean and renewable energy resources.[1] In fact, hydrogen ($H_2$) fuels production via photoelectrochemical (PEC) water splitting ($2H_2O + \hbar v \rightarrow 4H_2 + O_2$) represents one of the most challenging technologies for the production of clean carbon-neutral energy.[2] The water splitting process consists of both reduction and oxidation half-reactions, *i.e.*: i) oxygen evolution reaction (OER: $2H_2O \rightarrow O_2 + 4H^+ + 4e^-$) and ii) hydrogen evolution reaction (HER: $4H^+ + 4e^- \rightarrow 2H_2$).[3] In general, a water splitting PEC cell comprises a semiconductor photoelectrode and a counter electrode immersed in an aqueous electrolyte.[3-5] Semiconductor photoelectrodes absorb light photogenerating electrical charges, which are needed to carry out the redox chemistry of the OER and HER processes.[3-5] For a n-type semiconductor electrode (photoanode),[5] such as $TiO_2$[3,4] and $\alpha$-$Fe_2O_3$,[3] photoexcited holes are transferred to the semiconductor/electrolyte junction where they enable the OER,[3] while electrons are transferred to the counter electrode where they allow the HER.[3] Viceversa, in the case of a p-type semiconductor electrode (photocathode)[5] such as p-Si,[6] p-InP,[7a] and p-GaAs,[7b] HER and OER processes occur at the semiconductor/electrolyte junction and the counter electrode, respectively.[3-5] The solar to hydrogen conversion efficiency ($\eta_{STH}$) is the most important figure of merit (FoM) of a PCE cell and is defined as:

$$\eta_{STH} = \left| \frac{|j_{sc}| \times 1.23 \times \eta_F}{P} \right|_{AM1.5G}$$

where $j_{sc}$ is the short-circuit photocurrent density, $\eta_F$ the Faradaic efficiency for hydrogen evolution, and P the incident illumination power density, measured under standard solar illumination conditions (AM1.5G).[8] This FoM directly depends on the photophysical properties of the semiconductor photoelectrodes, such as light absorption,[4,5] exciton formation,[8] as well as charge carrier separation and transport.[8] The photogenerated electrons and holes have to overcome energetic constraints, corresponding to the thermodynamic potential of the HER and OER processes, respectively.[4,5,8] In particular, the electrochemical potential of the bottom of the photoelectrode conduction band must be more negative than the $H^+/H_2$ redox level ($E^0_{H^+/H_2}$ = 0 V),[3-5,8] while the one of the top of the valence band must be more positive than the $O_2/H_2O$ redox level ($E^0_{O_2H_2O}$ = 1.23 V).[3-8] Moreover, in view of scaling up and commercialization of PEC cells, stability and cost are also key factors.[2,3] Up to date, extensive research on photoelectrodes for PEC cells has been focused on inorganic metal oxide/nitride such as $TiO_2$,[3,4,9] $ZnO$,[10] $WO_3$,[11] $\alpha$-$Fe_2O_3$,[3,12] $BiVO_4$,[13] $Ta_2O_5$,[14] $TaON$[15] and $Ta_3N_5$,[15] due to their energy band gap exceeding 1.23 eV,[2-5,8-15] stability and Earth-abundance.[2,3,8] Nevertheless, these single semiconductor absorbers cannot harvest a significant portion of solar spectrum and therefore their potential $\eta_{STH}$ is intrinsically limited (predicted maximum values ~13%).[8-15] In order to overcome this problem, ideal PEC cell based on dual light absorber, or tandem, configuration, can be used. A simple two-photoelectrode approach to construct a PEC water splitting is to use an n-type semiconductor photoanode together with a p-type semiconductor photocathode.[8-9] In this context, the vertical stacking of a 1.6-1.8 eV energy band gap semiconductor, such as InP,[7] GaInP$_2$,[16] AlGaAs,[17] Cu$_2$O[18] on top of a narrower (~1 eV) one, such as Si,[6,17,19] is emerging as a promising route to optimize the solar photon harvesting (photons not absorbed by the first material are transmitted to and absorbed by the second one).[8-9,17,19] Recently, tandem water splitting using perovskite photovoltaics and CuIn$_x$Ga$_{1-x}$Se$_2$ photocathodes[20] or α-Fe$_2$O$_3$ photoanode[21] reached $\eta_{STH}$ of 6% and 2.4%, respectively. However, the instability in aqueous solutions of the photoactive semiconductors[6,7,16-21] is their main drawback for long time operation. To overcome this issue, encapsulation strategies of the photoactive semiconductors have been proposed,[18,22,23] with the result to raise the overall fabrication costs. Thus, the discover of new photoelectrode materials is needed to further improve the water splitting efficiency and long term stability respect to the current technology.[6,22] In the quest for the development of new photoelectrode materials, recently, organic semiconductors, such as graphitic carbon nitrides (g-C$_3$N$_4$),[24] 1,3,5-tris-(4-formyl-phenyl)triazine-based covalent organic frameworks (TFPT-COF),[25a] triarylamine-based microporous organic network (TAA-MON)[25b], hydrogen-bonded organic pigments of the epindolidione (EPI) and quinacridone (QNC),[26a] and boron subnaphthalocyanine chloride (SubPc):α-sexithiophene (6T) blend,[26b] have been exploited for the HER process. Sprick et al. reported conjugated microporous polymer (CMP) based on copolymer with varying ratios of benzene and pyrene as semiconductors for HER with optical band gap finely tuned over a broad range (*i.e.*, 1.94-2.95 eV), thus making them as versatile materials for tandem configurations.[27] Finally, PEC electrodes for HER based on semiconducting polymers (SPs) also emerged.[28,29] These studies focused mainly on the use of regio regular poly(3-hexylthiophene) (rr-P3HT),[28,29] the prototypical conjugated polymer widely used in bulk heterojunction (BHJ) organic photovoltaics (OPV) architectures,[30] capable of delivering a record photocurrent density (under 1 sun illumination) of up to 14 mA cm$^{-2}$.[31] rr-P3HT has a direct bandgap of 1.9 eV,[30,31] close to the optimum value for a PEC tandem device (maximum $\eta_{STH}$ of 21.6% is predicted stacking 1.89 eV and 1.34 eV energy band gap semiconductors).[8-9] The rr-P3HT lowest unoccupied molecular orbital (LUMO) energy level is several hundreds of millivolts more negative than the $E^0_{H^+/H_2}$ potential (LUMO$_{P3HT}$ - $E^0_{H^+/H_2}$ ~ -1,5 V),[30,31] thus photogenerated electrons possess the energy enabling the HER process.[32] Moreover, the optoelectronic properties of rr-P3HT, such as light absorption and charge photogeneration, are fully retained in aqueous environment.[33] On the basis of these observations, rr-P3HT:phenyl-C61-butyric acid methyl ester (PCBM) BHJ-based photocathodes with photocurrent densities at the RHE potential (J$_{0V vs RHE}$) in the mA cm$^{-2}$ range have been reported.[28-29,34] In these photocathodes, rr-P3HT:PCBM BHJ is sandwiched between two charge-selective layers (CSLs), resembling OPV architectures.[35] Specifically, the hole-selective layer (HSL) is deposited between a transparent conductive oxide (TCO), *e. g.* indium tin oxide (ITO)[32] or fluorine dope tin oxide (FTO),[28,29,32] and the rr-P3HT:PCBM,[28,29,32,34] while the electron-selective layer (ESL) is deposited on top of the rr-P3HT:PCBM.[28,29] The device is completed by depositing an electrocatalyst (EC) for the HER, giving the overall structure TCO/HSL/rr-P3HT:PCBM/ESL/EC.[28,29] The key role of the CSLs relies on the holes and electron transport[28,29,35] toward TCO and EC, respectively. Once the electrons are collected to the EC, the HER process is activated.[8-9,28,29] Currently, ZnO,[28a,34] $MoO_3$[29a,29b] and CuI[29d] are the most used HSLs, while $TiO_2$[28,29] and metallic Al/Ti[29c] are exploited as ESLs. Poly(3,4-ethylenedioxythiophene):poly(styrene sulfonate) (PEDOT:PSS),[36] the typical HSL in OPV,[37] poorly performed ($J_{0V\ vs\ RHE}$ in the sub mA cm$^{-2}$ region) in perovskites solar cells (PSCs),[29a] due to ion penetration and electrochemical doping processes, which could alter its electrical properties.[38] Amongst the ECs suitable for the HER process, both Pt[28a-d, 29b,d] and metal-free catalysts, such as $MoS_3$[28d,29a,c] and Chloro(pyridine)cobaloxime(III),[34] have been exploited, turning out photocathodes with $J_{0V\ vs\ RHE}$ above 8 mA cm$^{-2}$ and onset potential ($V_{oc}$) (defined as the potential at which the photocurrent related to the HER is observed) of ~0.7 V *vs.* RHE, a value similar to the open-circuit potential of the rr-P3HT:PCBM-based solar cell.[30]

Despite the continuous improvements, in terms of $J_{0V\ vs\ RHE}$ and $V_{oc}$, stability issues of the rr-P3HT:PCBM-based photocathodes have been evidenced for the CSLs, under $H_2$-evolving electrochemical conditions.[28,29] For instance, irreversible electrochemical degradation of hygroscopic PEDOT:PSS[28a,29c] and water-soluble CuI,[29d] intercalation processes and irreversible reduction into sub-stoichiometric phases of $MoO_3$[29b] and $WO_3$[28a] with unfavourable energy alignment for their use as HSL, leaded to corresponding photocathodes' lifetimes during continuous operation ranging from several minutes to few hours.[28a] Consequently, most attention is now focused toward the discover of new CSL materials, which could in principle boost the aforementioned photoelectrochemical FoMs.[39] In this context, two dimensional (2D)-transition metal dichalcogenides (TMDs) are raising interest, due to their optoelectronic properties, for the integration as CSL[40] in heterojunction-based solar cells,[41] both in OPV[40,41,43] and inorganic photovoltaics (IPV).[41,43,44] Amongst TMDs, $MoS_2$ is in a key position due to its high charge carriers mobility (up to ≈470 cm$^2$ V$^{-1}$ s$^{-1}$ for electrons, and ≈480 cm$^2$ V$^{-1}$ s$^{-1}$ for holes)[45] and chemical stability of the basal-planes.[45] In OPVs, solution-processed $MoS_2$ flakes have been exploited as HSL,[40,42,46,47] demonstrating power conversion efficiency (4% and 8% for P3HT:PCBM and PTB7:PCBM BHJs, respectively)[47] comparable to that of cells exploiting state-of-the-art HSLs, such as $MoO_3$[48] and PEDOT:PSS.[49] More recently, $MoS_2$ has also been exploited in PSCs,[50,51] as alternative hole transporting layer[50] to PEDOT:PSS[52] and Spiro-OMeTAD,[53] or as conducting and protecting buffer layer between the Spiro-OMeTAD and the perovskite layer.[51] Furthermore, the possibility to obtain thin $MoS_2$ flakes via chemical intercalation and liquid phase exfoliation of the bulk counterpart led to the formulation and application of functional $MoS_2$ inks in highly efficient, large-area solar cells fabricated by solution processing.[54]

In this work, we demonstrate the potentiality of 2D materials interface engineering by using few-layer $MoS_2$ flakes as HSL in organic PEC cells. In particular, we report the fabrication and photo-electrochemical study of solution-processed hybrid organic $H_2$-evolving photocathodes based on rr-P3HT:PCBM bulk heterojunction architecture sandwiched between solution-processed CSLs. $MoS_2$ flakes are used as HSL, while anatase $TiO_2$ nanoparticles act as ESL. $MoS_3$ nanoparticles, deposited on top of $TiO_2$, perform as Earth-abundant catalyst for the HER. The overall photocathode's structure is FTO/$MoS_2$/rr-P3HT:PCBM/$TiO_2$/$MoS_3$. Single- and few-layer $MoS_2$ flakes are obtained by using n-butyl lithium (n-BuLi) in cyclohexane as the intercalation agent to insert lithium ions into the layered bulk structure of $MoS_2$, followed by exfoliation in water with the aid of ultrasonication.[56] Solution-processed p-doping based on $HAuCl_4·3H_2O$ methanol solution[46,55] allows to tailor the Fermi levels, *i.e.* the work function (WF) values, of the $MoS_2$ films to higher values (from 4.6 eV of the pristine $MoS_2$ to 5.1 eV), thus matching the highest occupied molecular orbital (HOMO) level of rr-P3HT (~-5.1 eV).[30,31] Interface engineering allows to obtain uniform and fully-covered film morphology of $MoS_2$ flakes onto FTO, leading to solution-processed architectures with $J_{0V\ vs\ RHE}$ of 1.21 mA $cm^{-2}$, $V_{oc}$ of 0.55 V *vs.* RHE and power-saved FoM ($\Phi_{saved,NPAC}$) of 0.423%, thus approaching the state-of-the-art values of 0.47% for solution-processed rr-P3HT:PCBM-based photoacathodes,[29a] and of 1.21%[29d] obtained for non-encapsulated rr-P3HT:PCBM-based photocathodes (up to 1.45% and 2.05% in presence of PEI[29d] and Ti[29c] protective coatings, respectively).

## 2. Results and discussion
### 2.1. Characterization of solution processed $MoS_2$ flakes

$MoS_2$ flakes are prepared by a chemical lithium intercalation method,[57] avoiding the typical problems related to the LPE of $MoS_2$, *i.e.* the use of toxic and high boiling point solvents such as *N*-methyl-2-pyrrolidone (NMP) and dimethylformamide (DMF), typically used for the LPE of bulk $MoS_2$ and other layered materials,[54] as well as the presence of superficial oxide species[58] and high-temperature annealing processes needed for solvent removal.[59] Experimental details on the synthesis of the $MoS_2$ flakes are reported in the Experimental, Material preparation section.

The morphology of the as-produced $MoS_2$ flakes is characterized by means of transmission electron microscopy (TEM). Fig. 1a reports an image of the $MoS_2$ flakes, showing irregular shaped sheets. Statistical analysis of the TEM images, reported in Fig. 1b, indicates the presence of flakes with lateral size in the 30-800 nm range (mean value ~275 nm, *i.e.* mean area approx. 0.075 $μm^2$). Additional TEM images are reported in Supplementary Information (S.I.) (Fig. S1) to illustrate the morphology of $MoS_2$ flakes. Atomic force microscopy (AFM) analysis of the $MoS_2$ flakes deposited onto a V1-quality mica substrate is shown in Fig. 1c. A representative height profile of a single flake is also reported (red line in Fig. 1c), showing a thickness of 1.2 nm, while additional AFM images showing MoS$_2$ flakes with thickness up to 6 nm are reported in the S.I. (Fig. S2). Statistical analysis of the flakes thickness, reported in Fig. 1d, reveals the presence of single to few layers MoS$_2$ flakes, (monolayer thickness ~0.7-0.8 nm),[59] with average height values of 2.3 ± 1.6 nm.

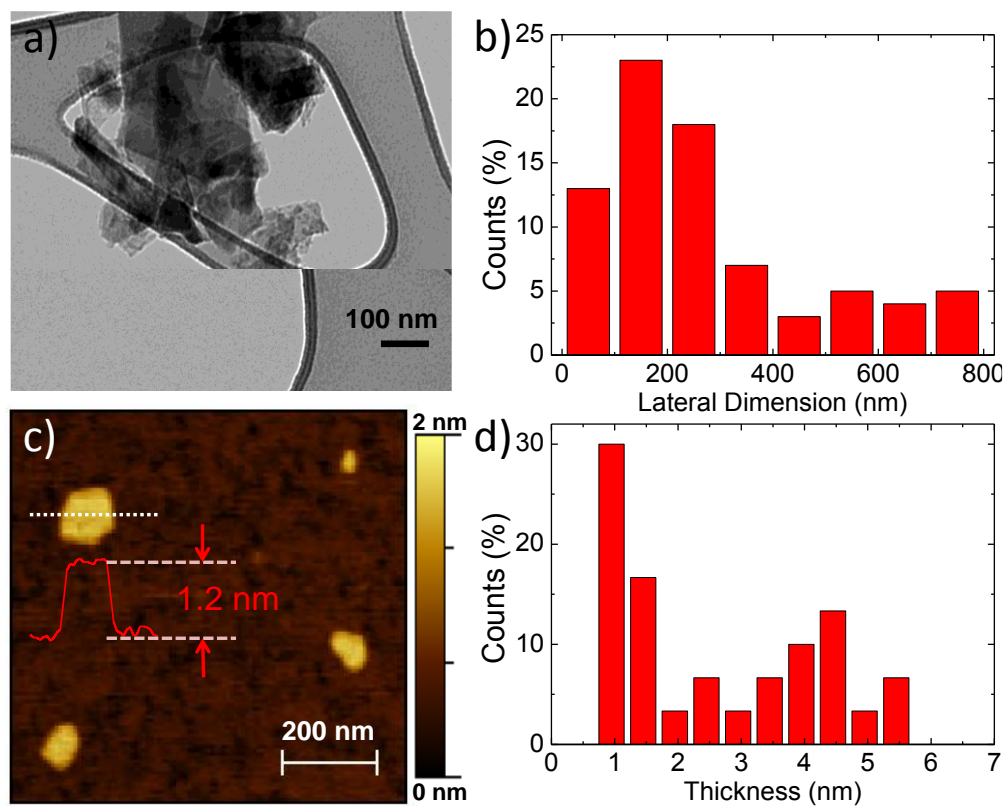

**Fig. 1** a) TEM images of the MoS$_2$ flakes and b) the statistical analysis of their lateral dimension (calculated on 80 flakes). c) AFM images of MoS$_2$ flakes deposited onto a V1-quality mica substrate. The height profile of a representative flake is also shown (red line). d) Statistical analysis of the flakes' thickness (derived from different AFM images and calculated on 30 flakes).

UV-Vis extinction spectra of MoS$_2$ flakes dispersed in isopropanol (IPA) is reported in Fig. 2a. Absorption peaks around 670 nm and 620 nm arise from the direct transitions from the valance band to the conduction band at the *K*-point of the Brillouin zone of layered MoS$_2$, known as the *A* and *B* transitions, respectively.[61] The broad absorption band centred at ~400 nm arises from the *C* and *D* inter-band transitions between the density of state peaks in the valence and conduction bands[62] of both 1*T* (metallic) and 2*H* (semiconducting) phases.[63]

Raman spectroscopy is carried out on the MoS$_2$ flakes, as well as on the bulk MoS$_2$, in order to confirm their different topological structure.[64,65] Representative spectra, reported in Fig. 2b, show the presence of first-order modes at the Brillouin zone center $E_{2g}^1(\Gamma)$ (~380 cm$^{-1}$ for MoS$_2$, ~378 cm$^{-1}$ for bulk MoS$_2$) and $A_{1g}(\Gamma)$ (~403 cm$^{-1}$ for both MoS$_2$ and bulk MoS$_2$), involving the in-plane displacement of Mo and S atoms and the out-of-plane displacement of S atoms, respectively.[64] The

$E_{2g}^1(\Gamma)$ mode of the MoS$_2$ flakes exhibits softening with respect to the one of the bulk counterpart, while no difference of the peak position of $A_{1g}(\Gamma)$ modes is observed. The shift of the $E_{2g}^1(\Gamma)$ mode is explained by dielectric screening of long range Coulomb interaction.[64,65] Thus, the frequency difference between $A_{1g}(\Gamma)$ and $E_{2g}^1(\Gamma)$ (24 cm$^{-1}$) of the MoS$_2$ decreases of about 2 cm$^{-1}$ with respect to the bulk MoS$_2$ (26 cm$^{-1}$).

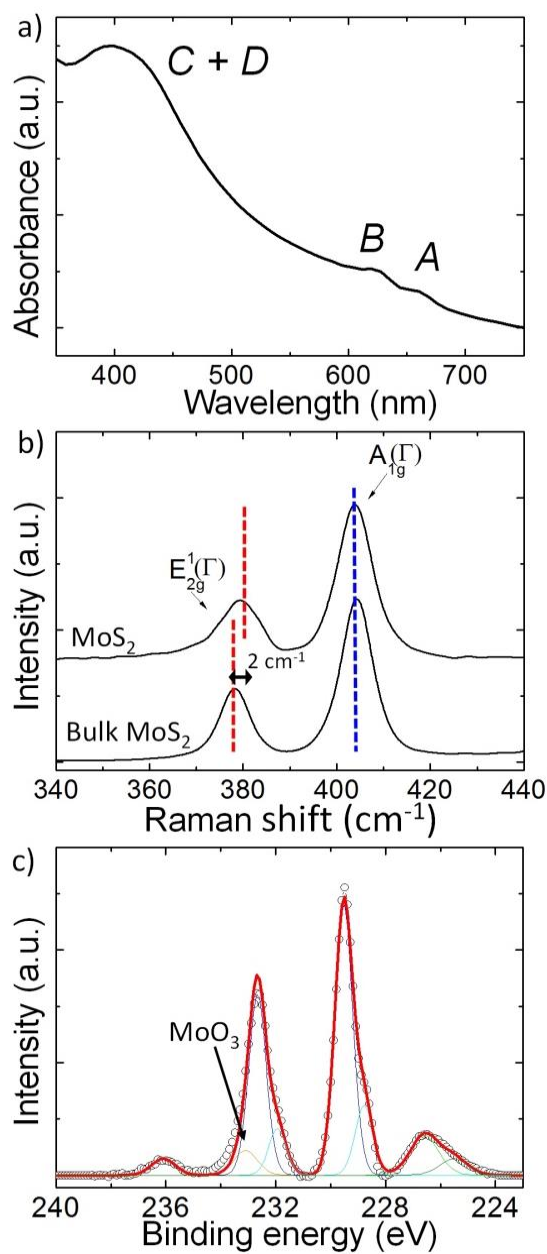

**Fig. 2** a) UV-Vis absorption spectra of 0.1 mg mL$^{-1}$ MoS$_2$ dispersion in IPA. *B* and *A* refer to the band at ~620 nm and ~670 nm, respectively, which arise from the direct transitions from the valance band to the conduction band at the *K*-point of the Brillouin zone. *C* + *D* band refers to the overlap of the inter-band transitions between the density of state peaks in the valence and conduction bands. b) Raman spectra of exfoliated MoS$_2$ flakes and bulk MoS$_2$ deposited onto a Si wafer with 300 nm thermally grown SiO$_2$. The main peaks located around 379 cm$^{-1}$ and 403 cm$^{-1}$ are attributed to the

$E_{2g}^1(\Gamma)$ and $A_{1g}(\Gamma)$ modes, respectively. Dashed vertical red line shows the frequency shift of $E_{2g}^1(\Gamma)$ mode of MoS$_2$ respect to that of bulk MoS$_2$. The dashed vertical blue line shows the same position of the $A_{1g}(\Gamma)$ modes of exfoliated MoS$_2$ and bulk MoS$_2$. c) XPS spectrum collected on the MoS$_2$ flakes casted onto 50 nm-Au sputtered coated Si wafers from the 0.1 mg mL$^{-1}$ dispersion in IPA.

The full width at half maximum (FWHM) of the $E_{2g}^1(\Gamma)$ and $A_{1g}(\Gamma)$ of MoS$_2$ increases of ~2 cm$^{-1}$ and ~1 cm$^{-1}$, respectively, compared to the corresponding bulk MoS$_2$ modes. In particular, the increase of $A_{1g}(\Gamma)$ FWHM for MoS$_2$ has been linked with the variation of interlayer force constants between the inner and outer layers.[66] Statistical Raman analysis of the MoS$_2$ is reported in S.I. (Fig. S3).

XPS measurements are carried out on the MoS$_2$ flakes to determine their elemental composition, *i.e.* their chemical quality resulting from the preparation by chemical lithium intercalation method.[57,58] The XPS spectrum, reported in Fig. 2c, can be de-convoluted by four peaks: the first two peaks (fitted with two components) centred at 226 eV and 229 eV are assigned to S$_{2s}$ and Mo$_{3d}$ of MoS$_2$, respectively; the third peak is also assigned to Mo$_{3d}$ and it is fitted with three components. The latter component (peaked at 233 eV) and the fourth peak centred at 236 eV are attributed to the MoO$_3$ phase, which are produced as by-product for MoS$_2$ flakes exfoliated and exposed to air.[58] However, we remark that the MoO$_3$-related peaks in our Li-exfoliated flakes are significantly reduced with respect to those observed in MoS$_2$ flakes produced by LPE in NMP,[51] whose XPS spectrum has been previously reported also by our group,[58] thus confirming the high quality of the MoS$_2$ flakes produced in this work.

The effects on the surface morphology of the FTO after the MoS$_2$ flakes deposition is microscopically investigated by AFM. Fig. 3a reports the AFM image of the bare FTO, while in Fig. 3b it is shown the one of the FTO/MoS$_2$. FTO/MoS$_2$ shows nano-step height modulations on the grained FTO (grain size >100 nm).[62] Representative height profiles of the AFM images are reported in Fig 3c and Fig. 3d for FTO and FTO/MoS$_2$, respectively. For the case of FTO/MoS$_2$ the edge steps of zoomed height profile (as defined by the blue dashed rectangular) are in the 1-1.5 nm range, in agreement with the flakes thickness (2.3 ± 1.6 nm) measured by AFM (Fig. 1c-d).

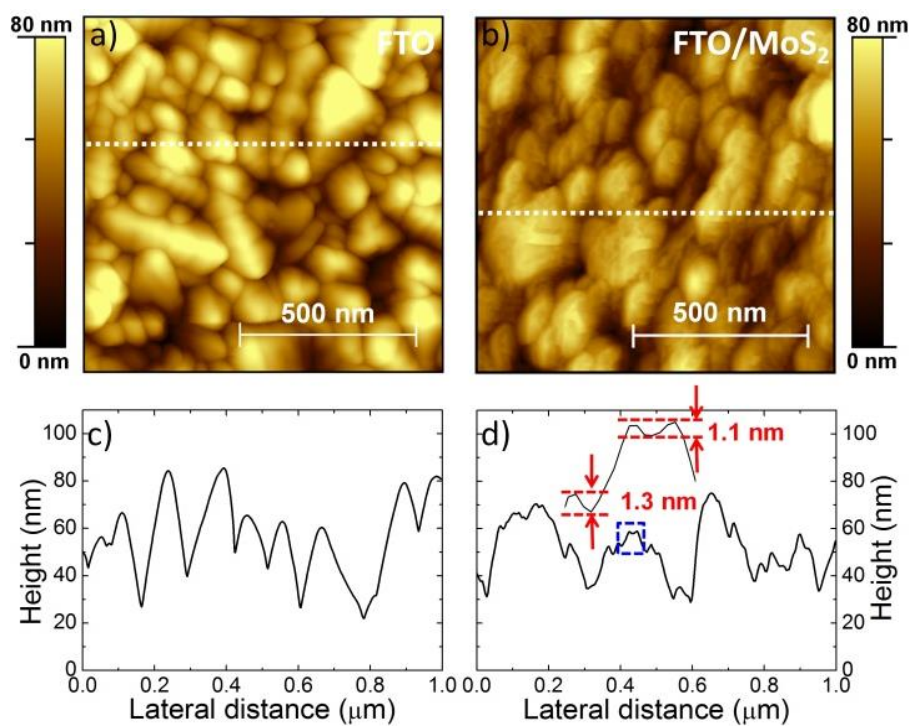

**Fig. 3** a) AFM images of FTO and b) FTO/MoS$_2$. The MoS$_2$ films are deposited from a 0.1 mg mL$^{-1}$ MoS$_2$ dispersion in IPA. Representative height profiles of c) FTO and d) FTO/MoS$_2$ (the correspondent profile positions are shown by the dashed lines in a) and b), respectively). In d) the zoom of the height profile defined from the blue dashed rectangular, showing nano-edge steps, is reported.

MoS$_2$ flakes deposited onto FTO, see experimental, are subsequently doped by spin coating HAuCl$_4$·3H$_2$O methanol solution on top of the flakes themselves.[46,55] The doping process is a consequence of the positive reduction potential of the HAuCl$_4$, which is then prone to accept electrons from MoS$_2$ carrying out the reduction of Au$^{3+}$ to Au$^0$ species.[46,55] As a consequence, the electrical properties of the MoS$_2$ film, i.e., conductivity and WF value, are significantly affected. The doping level is modulated by varying the concentration of HAuCl$_4$·3H$_2$O solutions. Values of concentration of 5, 10, and 20 mM are tested (see details in Experimental, Fabrication techniques), giving the MoS$_2$-based films here named as p-MoS$_2$ (5 mM), p-MoS$_2$ (10 mM) and p-MoS$_2$ (20 mM), respectively. Similar dopant concentrations have been already used for engineering the WF of solution-processed MoS$_2$ thin-films for hole transport layers in OPVs.[46] The WF value of MoS$_2$, as measured by Kelvin Probe (KP) in ambient conditions (Table 1), is 4.6 eV, thus similar to the one measured for FTO (4.7 eV). After doping, the WF values of the p-MoS$_2$ (5 mM) (4.9 eV) and p-MoS$_2$ (10 mM) (5.1 eV) increase by 0.3 and 0.5 eV, respectively, compared to the one shown by the pristine MoS$_2$ film. Further increase of the doping level up to 20 mM does not reflect an additional rise of the WF value.

**Table 1** WF values of FTO, MoS$_2$ and p-MoS$_2$ (5, 10, 20 mM), as measured by ambient KP.

| MATERIAL | WORK FUNCTION (eV) |
|---|---|
| FTO | 4.7 |
| MoS$_2$ | 4.6 |
| p-MoS$_2$ (5 mM) | 4.9 |
| p-MoS$_2$ (10 mM) | 5.1 |
| p-MoS$_2$ (20 mM) | 5.1 |

Fig. S4 reports the AFM images of the FTO/p-MoS$_2$ (10 mM), which shows no differences in surface morphology with respect to the un-doped case (FTO/MoS$_2$). The roughness average (Ra) values are reported in Table 2, showing a decrease of about 2 nm for both FTO/MoS$_2$ and FTO/p-MoS$_2$ (10 mM) (Ra values of 11.6 nm and 11.9 nm, respectively) if compared with the value of the bare FTO (Ra = 13.8). Thus, the FTO roughness is reduced by the overlayer of MoS$_2$ flakes, which could be linked with their planarity and face-on arrangement.

**Table 2** Roughness average (Ra) values of the FTO, FTO/MoS$_2$ and FTO/p-MoS$_2$ (10 mM). The MoS$_2$ films are deposited from a 0.1 mg mL$^{-1}$ MoS$_2$ dispersion in IPA.

| SAMPLE | Roughness average (Ra) (nm) |
|---|---|
| FTO | 13.8 |
| FTO/MoS$_2$ | 11.6 |
| FTO/p-MoS$_2$ (10 mM) | 11.9 |

Fig. 4 reports the top view scanning electron microscopy (SEM) images of the FTO, FTO/MoS$_2$, FTO/p-MoS$_2$ (10 mM) and FTO/p-MoS$_2$ (20 mM) samples. No modifications of the FTO surface are observed after the MoS$_2$ flakes deposition and doping treatment, *i.e.*, 10 mM HAuCl$_4$·3H$_2$O, in agreement with the AFM data, i.e., Ra values reported in Table 2. The increase of doping level to 20 mM HAuCl$_4$·3H$_2$O determines the formation of some aggregates, thus affecting the surface's homogeneity, as revealed by Fig. 4d. These aggregates are attributed to the precipitation of Au and MoO$_3$ clusters after Au ions reduction and Mo$^{4+}$ to Mo$^{6+}$ conversion processes.[46,55] In fact, for doping level exceeding 10 mM, more electrons are needed to reduce the increased number of Au ions, and thus Mo$^{4+}$ can be converted into Mo$^{6+}$, resulting in the formation of Au nanoparticles and MoO$_3$.[46,55]

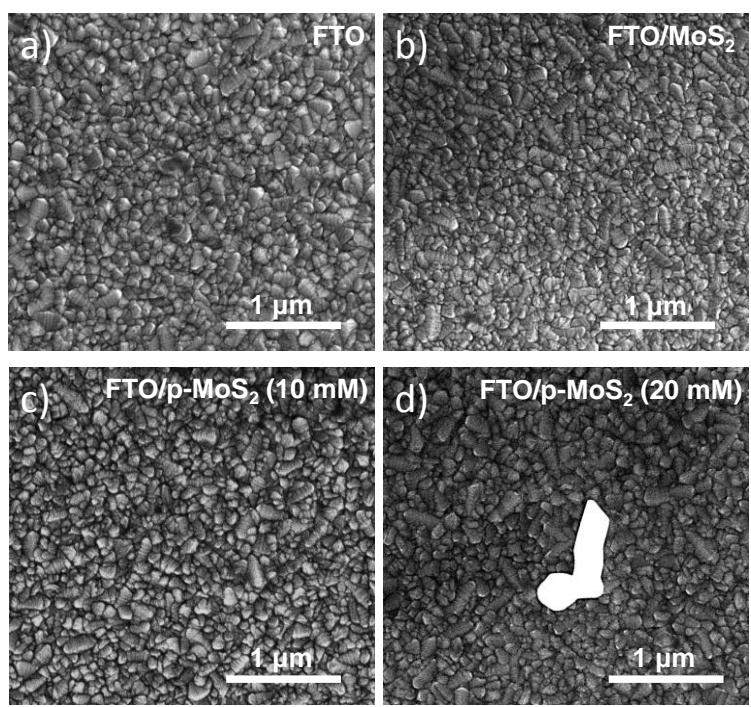

**Fig. 4** Top view SEM images of the a) bare FTO, and b) FTO/MoS$_2$, c) FTO/p-MoS$_2$ (10 mM) and d) FTO/p-MoS$_2$ (20 mM) films. The MoS$_2$-based films are deposited from a 0.1 mg mL$^{-1}$ MoS$_2$ dispersion in IPA.

### 2.2. Architecture of the hybrid solution-processed organic H$_2$-evolving photocathode

The full structure of the solution-processed hybrid organic H$_2$-evolving photocathode consists of a photoactive rr-P3HT:PCBM BHJ sandwiched between solution-processed CSLs. Few-layer MoS$_2$ flakes are spin coated onto FTO and tested as HSL, while anatase TiO$_2$ nanoparticles are spin coated onto rr-P3HT:PCBM and used as ESL. Amorphous MoS$_3$ nanoparticles, deposited on top of the TiO$_2$, act as EC for HER. Additional details of the fabrication of the photocathodes are reported in Experimental, Fabrication techniques.

Fig. 5a shows the representative energy band edge positions of the semiconductors of the hybrid photocathode together with the redox levels of the HER and OER. MoS$_2$, as HSL, is expected to extract the photogenerated holes towards the back conductive substrate (FTO) while the TiO$_2$ (as ESL) transports the photogenerated electrons towards MoS$_3$ (EC). Here, aqueous protons are reduced to H2, which evolves from the photocathode surface.[28,29] In order to provide the electrical driving force for the holes' collection,[28,29] MoS$_2$ films are doped by spin coating HAuCl$_4$·3H$_2$O methanol solutions on top of them,[46,55] thus increasing the WF values of the films from 4.6 eV up to 5.1 eV (for p-MoS$_2$ (10 mM) and p-MoS$_2$ (20 mM), as previously discussed (see also Table 1). Fig. 5b shows the high-resolution cross-sectional SEM image of a representative photocathode FTO/p-MoS$_2$ (10 mM)/rr-P3HT:PCBM/TiO$_2$/MoS$_3$. The layers are consecutively deposited by spin coating (see details in Experimental, Fabrication techniques), giving a well-defined multi-layered structure.

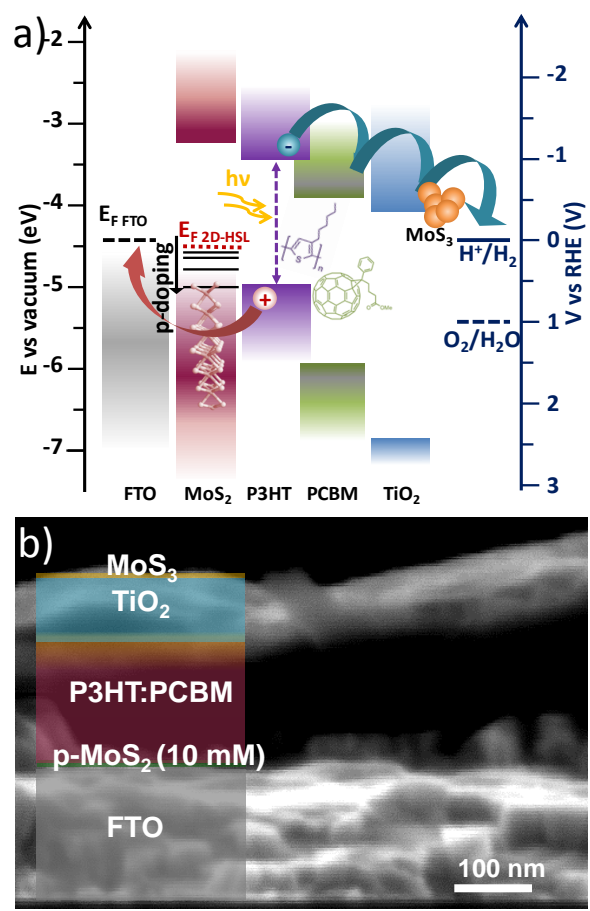

**Fig. 5** a) Typical energy band edges position of materials assembled in the hybrid solution-processed photocathode. The rr-P3HT:PCBM layer, in BHJ configuration, efficiently absorbs light and generates charges. $MoS_2$ and $TiO_2$ act as hole- and electron- selective layers (HSL and ESL), respectively, driving the holes towards the FTO substrates and the electrons towards the $MoS_3$ nanoparticles, acting as EC layer for HER. Redox levels of both hydrogen evolution reaction (HER) (blue solid line) and oxygen evolution reaction (OER) (blue dashed line) are also shown. WF values of the $MoS_2$ (4.6 eV) is measured by ambient KP. These values can be tailored to higher values (up to 5.1 eV) by solution-processed treatments using $HAuCl_4·3H_2O$ as dopant agent. b) High-resolution cross-sectional SEM image of the representative photocathode FTO/p-$MoS_2$ (10 mM)/rr-P3HT:PCBM/$TiO_2$/$MoS_3$.

$MoS_2$ and $MoS_3$ layers are hard to be resolved because their thickness is small with respect to the device-scale magnification. Thicknesses around 150 and 80 nm can be estimated for the rr-P3HT:PCBM and $TiO_2$ layers, respectively. The same thickness values of the rr-P3HT:PCBM layer is also measured by profilometry (see Experimental, Fabrication techniques).

### 2.3. Photoelectrochemical characterization

The rrP3HT:PCBM-based photocathodes using $MoS_2$ and p-$MoS_2$ as HSLs are characterized by linear sweep voltammetry (LSV) in $H_2SO_4$ solution at pH 1. The results for $MoS_2$ and p-$MoS_2$ (10 mM) deposited from 0.1 mg mL$^{-1}$ $MoS_2$ dispersion in IPA are reported in Fig. 6a, where they are compared with the responses of a HSL-free photocathode and the current-potential curve of $MoS_3$ EC (deposited directly onto FTO). $MoS_3$ EC reveals activity for HER, with onset overpotential of 180 mV with respect to the RHE potential.[63] The voltammograms of the photocathodes show a photocurrent that increases when the potential is swept towards negative values. The photocurrents are positively affected by the presence of $MoS_2$ films, thus confirming their role as HSL. The dependence on the doping level of the p-$MoS_2$-based photocathodes is shown in Fig. 6b. It is worth to note that different concentrations (0.05, 0.1, 0.4 mg mL$^{-1}$) of $MoS_2$ dispersion in IPA have been also preliminary tested without doping (Fig. S5) and with 10 mM doping (Fig. S6), showing the best results for the 0.1 mg mL$^{-1}$ dispersion.

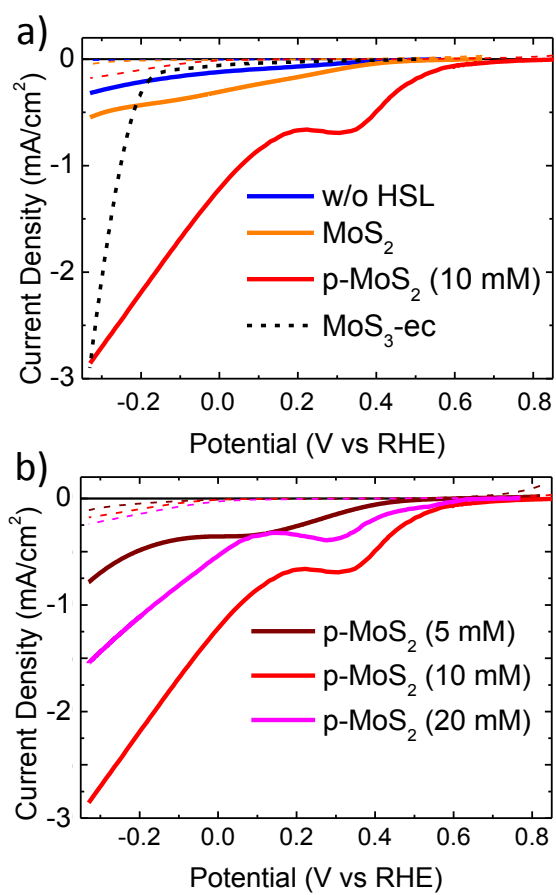

**Fig. 6** a) LSVs measured for the photocathodes using $MoS_2$ (orange lines) and p-$MoS_2$ (10 mM) (red lines) as HSLs. measured in 0.5 M $H_2SO_4$ solution at pH 1, under dark (dashed lines) and AM1.5 light illumination (100 mW cm$^{-2}$) (solid lines) $MoS_2$ films are deposited by 0.1 mg mL$^{-1}$ $MoS_2$ dispersion in IPA. The photoelectrochemical responses of the photocathode without any HSL (blue lines) and the current-potential curve of $MoS_3$ electrocatalyst (deposited directly on FTO) (short dashed black line) are also shown. b) LSVs measured for p-$MoS_2$-based photocathodes obtained with different level of doping (purple, red and magenta lines for 5, 10 and 20 mM $HAuCl_4·3H_2O$ methanol solution, respectively), in the same previous conditions.

The photovoltage $V_{photo}$ is the difference between the potential applied to the photocathode under illumination ($E_{light}$) and the potential applied to the catalyst electrode ($E_{dark}$) to obtain the same current density. The subscript "m" stands for "maximum". The power-saved FoM ($\Phi_{saved,NPAC}$) reflects the photovoltage and photocurrent of a photocathode independently from the over-potential requirement of the catalyst.[64] Here, we simply assume that the MoS$_3$ film deposited onto FTO is identical to the one deposited onto TiO$_2$. Table 3 summarizes the main FoMs extracted from the voltammograms measured for the different cases: the photocurrent density taken at 0 V *vs.* RHE (J$_{0V\ vs\ RHE}$), the V$_{oc}$, and $\Phi_{saved,NPA,C}$, which is calculated by:

$$\Phi_{saved,NPAC} = \eta_F \times \frac{|j_{photo,m}| \times [E_{light}(J_{photo,m}) - E_{dark}(J_{photo,m})]}{P_{in}}$$

$$= \eta_F \times \frac{|j_{photo,m}| \times V_{photo,m}}{P_{in}}$$

where $\eta_F$ is the Faradaic efficiency assumed to be 100 %,[28d, 29a, 29c] $P_{in}$ is the power of the incident illumination and $j_{photo,m}$ and $V_{photo,m}$ are the photocurrent and photovoltage at the maximum power point, respectively. $j_{photo,m}$ is obtained by calculating the difference between the current under illumination of a photocathode and the current of the corresponding catalyst.

**Table 3** FoMs of photocathodes fabricated without HSL and with MoS$_2$ and p-MoS$_2$ (5, 10 and 20 mM) as HSL: the current density taken at 0 V vs. RHE (J$_{0V\ vs\ RHE}$), the onset potential (V$_{oc}$), defined as the potential at which a photocurrent density of 0.1 mA cm$^{-2}$ is reached, and the power-saved FoM $\Phi_{saved,NPA,C}$.

| HSL | J$_{0V\ vs\ RHE}$ (mA cm$^{-2}$) | V$_{oc}$ (V vs. RHE) | $\Phi_{saved,NPAC}$ (%) |
|---|---|---|---|
| - | 0.12 | 0.07 | 0.015 |
| MoS$_2$ | 0.30 | 0.30 | 0.070 |
| p-MoS$_2$ (5 mM) | 0.36 | 0.35 | 0.095 |
| p-MoS$_2$ (10 mM) | 1.21 | 0.55 | 0.423 |
| p-MoS$_2$ (20 mM) | 0.54 | 0.49 | 0.192 |

Although the value of the main FoMs of the undoped MoS$_2$-based photocathode ($J_{0V\ vs\ RHE}$ = 0.3 mA cm$^{-2}$, $V_{oc}$ = 0.3 V vs. RHE, $\Phi_{saved,NPAC}$ = 0.070%) increases with respect to the ones of the HSL-free photocathode ($J_{0V\ vs\ RHE}$ = 0.12 mA cm$^{-2}$, $V_{oc}$ = 0.07 V vs. RHE, $\Phi_{saved,NPAC}$ = 0.015%), significant photocurrents are observed only using p-MoS$_2$ (above 1 mA/cm$^2$ for potential <0.06 V vs. RHE for a doping level of 10 mM). For p-MoS$_2$ (10 mM) and p-MoS$_2$ (20 mM), the presence of a photoreduction peak, located at ~0.33 V vs. RHE, is also observed. Its origin is attributed to the chemical reduction of MoS$_3$ EC towards more electrocatalytically active MoS$_{2+x}$ species for HER,[68] as also confirmed by the analysis of two consecutive LSVs of FTO/MoS$_3$ measured in the same testing conditions as the ones reported in Fig. 6 (see S.I. Fig. S7). In the case of p-MoS$_2$ (10 mM), $J_{0V\ vs\ RHE}$, $V_{oc}$ and $\Phi_{saved,NPAC}$ are 1.21 mA cm$^{-2}$, 0.56 V *vs.* RHE and 0.423%, respectively. Photocathodes based on p-MoS$_2$ (20 mM) (highest doping level) and p-MoS$_2$ (5 mM) (lowest doping level) report $J_{0V\ vs\ RHE}$ of 0.54 and 0.36, thus decreasing by 56% and 70% with respect to the case of p-MoS$_2$ (10 mM), respectively. Concerning p-MoS$_2$ (5 mM), also in this case there is a decrease of the $V_{oc}$ ($V_{oc}$= 0.35 V *vs.* RHE) of 200 mV with respect to the values observed for p-MoS$_2$ (10 mM). Moreover, we noted a decrease of 60 mV of the $V_{oc}$ for the p-MoS$_2$ (20 mM) if compared with the p-MoS$_2$ (10 mM) photocathode. The values of $\Phi_{saved,NPAC}$ calculated for p-MoS$_2$ (5 mM) and p-MoS$_2$ (20 mM) are 0.095% and 0.192%, respectively. These values correspond to a decrease of 77.5% and 54.6% with respect to that of p-MoS$_2$ (10 mM), respectively.

The obtained results highlight the importance of the WF tuning of the MoS$_2$ films by p-doping treatment for their full-exploitation as highly performant HSLs. Fig. 7 shows the MoS$_2$ doping level dependence of the photocathodes' FoMs, as gathered from Table 3. In agreement with the SEM characterization reported in Fig. 4d, the decrease of the $J_{0V\ vs\ RHE}$ for the p-MoS$_2$ (20 mM) case could be ascribed to charge recombination pathways, *i.e.*, leakage currents, in presence of blend-uncovered Au and MoO$_3$ clusters. Although these surface alterations of the MoS$_2$ films negatively affect the photocurrents, the $V_{oc}$ value is similar to the one recorded for p-MoS$_2$ (10 mM). Differently, for the case of p-MoS$_2$ (5 mM), we observed a decrease of both the $J_{0V\ vs\ RHE}$ and the $V_{oc}$ with respect to the p-MoS$_2$ (10 mM) case. This result could be linked with differences (homogeneity) of the MoS$_2$ film doping, as suggested by the lower WF value of p-MoS$_2$ (5 mM) film (4.9 eV) if compared with the ones of p-MoS$_2$ (10 mM) and p-MoS$_2$ (20 mM) films (5.1 eV).

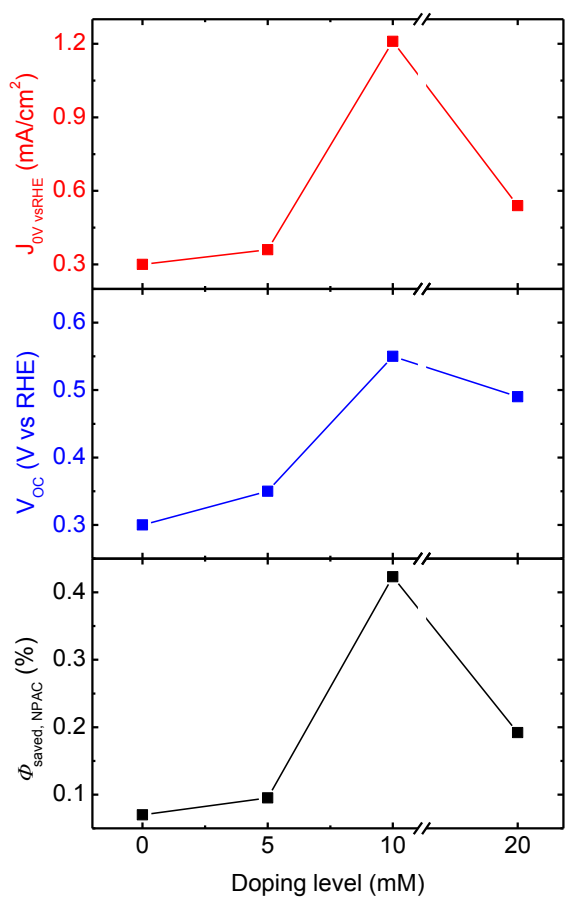

**Fig. 7** Dependence of the photocatodes' FoMs (red, blue and black colours for $J_{0V\ vs\ RHE}$, $V_{oc}$ and $\Phi_{saved,NPAC}$, respectively) on the HAuCl$_4$·3H$_2$O doping level (from 0 to 20 mM) of the p-MoS$_2$ films.

As a consequence, the main characteristics of the p-MoS$_2$ (5 mM)-based photocathode ($V_{oc}$ ~0.35 V *vs.* RHE, $J_{0V\ vs\ RHE}$ ~0.36 mA cm$^{-1}$ and $\Phi_{saved,NPAC}$ ~0.095) are similar to those recorded for the undoped MoS$_2$-based photocathode ($V_{oc}$ ~0.30 V *vs.* RHE, $J_{0V\ vs\ RHE}$ ~0.30 mA cm$^{-1}$ and $\Phi_{saved,NPAC}$ ~0.070). To sum up, $V_{oc}$ seems to be linearly correlated with the WF values of the different films, while the $J_{0V\ vs\ RHE}$ turns out also to be affected by the films surface morphology. The combination of these effects explain the behaviour of the $\Phi_{saved,NPAC}$, whose best values of 0.423% is reached for the p-MoS$_2$ (10 mM)-based photocathode.

Stability test under potentiostatic condition is performed for p-MoS$_2$ (10 mM) case, recording in time the $J_{0V\ vs\ RHE}$ under continuous illumination (Fig. 8). The data show an average initial photocurrent value of 1.36 mA cm$^{-2}$ followed by a steep decrease in performance down to 0.77 mA cm$^{-2}$ during the first 5 minutes (photocurrent loss of 50.7%). After 30 minutes of continuous operation, the photocurrent density reaches 0.49 mA cm$^{-2}$ (photocurrent loss of 63.2%). Thus, the photocurrent mainly decreases at the beginning of the illumination process. The quick, initial performance degradation has been recently observed in similar architectures using Pt as catalyst and CuI as HSL,[29d] being attributed to the irreversible Pt detachment from the surface of the electrodes.[29d] A similar effect could affect the MoS$_3$ EC used here. After more than 5 min, a progressive stabilization is observed. Notably, there is no delamination of the film during the measurements, suggesting the electrochemical stability of the FTO/p-MoS$_2$/P3HT:PCBM underlayers. Moreover, cyclic voltammetry (CV) measurements reveal the absence of irreversible redox reactions involving MoS$_2$ films under the HER-working condition of the photocathodes (see S.I., Fig. S8). This is in agreement with the electrochemical stability of the MoS$_2$[45] expressed in several energy conversion and storage devices,[70] including fuel and water splitting cells,[71] batteries[72] and supercapacitors.[73] These results evidence the need of future efforts also in the development of stable ESL/EC layers, for which the integration of TMDs could offer winning solutions thanks to their tuneable charge-selective properties[46] and electro-catalytic activity towards HER.[70]

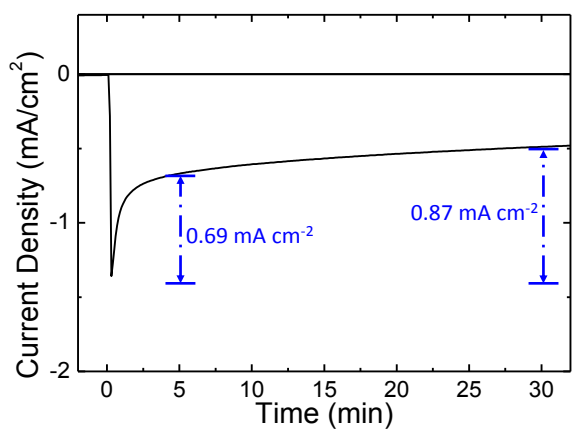

**Fig. 8** Potentiostatic stability test of photocathode fabricated with p-MoS$_2$ (10 mM) at pH 1 and 0 V *vs.* RHE under continuous illumination (1 sun). The blue lines mark the losses of the photocurrent after 5 and 30 min.

## 3. Conclusion

Few-layer flakes MoS$_2$ are demonstrated as novel HSL for solution-processed hybrid organic H$_2$-evolving photocathodes. The production method of the MoS$_2$ flakes underpins on water-based exfoliation of Li-intercalated bulk MoS$_2$, avoiding the use of toxic reagents/solvents and being potentially compatible with large-area and low-cost production. We produced ultrathin MoS$_2$ layer (<10 nm) that, once deposited, not affect the morphology of the underneath FTO substrate. We tuned the electrical properties of the MoS$_2$ films by using solution-processed HAuCl$_4$·3H$_2$O doping, increasing the WF value from 4.6 up to 5.1 eV, thus well matching with the HOMO level of the rr-P3HT. The morphology of the MoS$_2$ films and the optimization of the HAuCl$_4$·3H$_2$O doping level led to a hybrid solution-processed organic H$_2$-evolving photocathodes with J$_{0V\ vs\ RHE}$ of 1.21 mA cm$^{-2}$ at 0 V *vs.* RHE, V$_{oc}$ of 0.56 V *vs.* RHE and $\Phi_{saved,NPA,C}$ approaching 0.5%. These results pave the way towards the integration of MoS$_2$ as HSL in organic PEC cells to boost this technology as efficient scalable and low-cost tool for solar H$_2$ production.

## 4. Experimental

### 4.1. Material preparation

MoS$_2$ flakes are prepared by a chemical lithium intercalation method. Experimentally, 0.3 g of MoS$_2$ bulk powder (Sigma Aldrich) is dispersed in 4 mL of 2.0 M n-butyllithium (n-BuLi) in cyclohexane[56] (Sigma Aldrich). The dispersion is kept stirring for two days at room temperature under Ar atmosphere. The Li-intercalated material ($Li_xMoS_2$) is separated by suction filtration under Ar. $Li_xMoS_2$ is washed with anhydrous hexane to remove non-intercalated Li ions and organic residues. $Li_xMoS_2$ powder is then exfoliated by ultrasonication (Branson ®5800 ultrasonic cleaner) in deionized (DI) water for 1 h. The obtained dispersion is then ultracentrifugated (Optima™ XE-90 ultracentrifuge, Beckman Coulter) at 10K for 20 min to remove LiOH and un-exfoliated material. Finally, the precipitate is filtered and re-dispersed in IPA (absolute alcohol, without additive, ≥ 99.8%, Sigma Aldrich), in order to accurately control the concentration of the final dispersions.

### 4.2. Fabrication techniques

Photocathodes are fabricated according to the architecture FTO/HSL/rr-P3HT:PCBM/$TiO_2$/$MoS_3$, where few-layer $MoS_2$ flakes are used as HSL. Complementary architectures without HSL are also fabricated. FTO coated soda-lime glass substrates (Dyesol, sheet resistance 15 Ω $sq^{-1}$) are cleaned according to the following protocols: sequential sonication baths in DI water, acetone, IPA each lasting for 10 minutes and plasma cleaning in an inductively coupled reactor for 20 minutes (100 W RF power, excitation frequency 13.56 MHz, 40 Pa of $O_2$ gas process pressure, background gas pressure 0.2 Pa).

$MoS_2$ dispersion is deposited onto the previously treated FTO by spin coating (Laurell Tech. Corp. Spin coater) using a single step spinning protocol with rotation speed of 3000 rpm for 60 s. Concentration of the dispersions of 0.05, 0.1 and 0.4 mg $mL^{-1}$ are tested. Post thermal annealing in Ar atmosphere at 150 °C for 30 min is performed for the $MoS_2$ films. The latter are subsequently doped by spin casting $HAuCl_4·3H_2O$ (≥99.9% trace metals basis, Sigma Aldrich) in methanol (ACS reagent, ≥99.8%, Sigma Aldrich) solution as p-doping agents on top, by using the same single step spinning protocol of the $MoS_2$ deposition. Doping concentrations of 5, 10, and 20 mM are tested. All doping solutions are sonicated for 10 minutes before deposition. The doped films are subsequently dried for 30 min under Ar atmosphere. The organic polymer film used in all the architectures consists in a blend of rr-P3HT, as the donor component, and PCBM, as the acceptor component (rr-P3HT:PCBM). rr-P3HT (electronic grade $M_n$ 15000-45000, Sigma Aldrich) and PCBM (99.5% purity, Nano C) are separately dissolved in chlorobenzene (ACS grade, Sigma Aldrich), at a 1:1 wt ratio and 25 mg $mL^{-1}$ on a polymer basis. Polymer blend solution is stirred at 40 °C for 24 hours before use. Blend thin films are obtained by spin casting the rr-P3HT:PCBM solution using the following set of parameters: two step spinning protocol with rotation speeds of 800 rpm for 3 s followed by 1600 rpm for 60 s, respectively. This spin casting protocol produced a rr-P3HT:PCBM blend layer 200 ± 20 nm thick, as measured by means of a Dektak XT profilometer (Bruker) equipped with a diamond-tipped stylus (2 mm) selecting a vertical scan range of 25 mm with 8 nm resolution and a stylus force of 1 mN, on an area of 0.25 $cm^2$. The sol-gel procedure for producing $TiO_2$ dispersion has been previously reported by Kim et al.[74] Thus, $TiO_2$ precursor solution is prepared in IPA and subsequently deposited by spin casting on top of the rr-P3HT:PCBM film as ESL. Subsequently, during 12 h in air at room temperature, the precursor converted to $TiO_2$ by hydrolysis. A three step spinning protocol with rotation speeds of 200 rpm for 3 s, 1000 rpm for 60 s and 5000 rpm for 30 s is used. Post thermal annealing in an Ar atmosphere is carried out at 130 °C for 10 min for all the devices before catalyst deposition. The devices are completed by a layer of $MoS_3$ nanoparticles (Alfa Aeasar) acting as catalyst for the HER process. The catalyst layer is obtained by spin casting a 3.8 mg mL$^{-1}$ water:acetone:NaOH (1M) 1:2:0.2 dispersion on top of the $TiO_2$. The dispersion is stirred overnight at room temperature and sonicated for 10 minutes before its use. Spinning protocol is identical to the one adopted for the $TiO_2$ deposition.

### 4.3. Material characterization

UV-Vis absorption spectrum of the 0.1 mg mL$^{-1}$ $MoS_2$ dispersion in IPA is obtained using a Cary Varian 5000UV-Vis spectrometer.

Raman measurements are carried out with a Renishaw 1000 using a 50X objective, a laser with a wavelength of 532 nm and an incident power of 1 mW. The different peaks are fitted with Lorentzian functions. For each sample 30 spectra are collected. 0.01 mg mL$^{-1}$ $MoS_2$ dispersion in IPA is drop-casted onto Si/$SiO_2$ (300 nm $SiO_2$) substrate and dried under vacuum. Statistical analysis of the relevant features is carried out by means of software Origin 8.1 (OriginLab).

The XPS analysis is carried out using a Kratos Axis Ultra spectrometer on $MoS_2$ flakes drop casted onto 50 nm-Au sputtered coated Si wafers from the 0.1 mg mL$^{-1}$ dispersion in IPA. The XPS spectra are acquired using a monochromatic Al Kα source operated at 20 mA and 15 kV. High-resolution spectra are collected with pass energy of 10 eV and energy step of 0.1 eV over a 300 μm x 700 μm area. The Kratos charge neutralizer system is used on all specimens. Spectra are charge corrected to the main line of the $C_{1s}$ spectrum set to 284.8 eV, and analyzed with CasaXPS software (version 2.3.17).

Transmission electron microscopy images are taken on a JEM 1011 (JEOL) transmission electron microscope, operating at 100 kV. 0.01 mg mL$^{-1}$ $MoS_2$ dispersions in IPA are drop-casted onto carbon coated Cu TEM grids (300 mesh), rinsed with DI water and subsequently dried under vacuum overnight. Lateral dimension of the $MoS_2$ flakes is measured using ImageJ software (Java). Statistical TEM analysis is carried out by means of software Origin 8.1 (OriginLab).

Scanning electron microscopy analysis is carried out with a FEI Helios Nano lab field-emission scanning electron microscope. The samples are imaged without any metal coating or pre-treatment. The samples are prepared depositing $MoS_2$ films on FTO as described in the Fabrication techniques section.

Atomic force microscopy images are obtained using commercial AFM instrument MFP-3D (Asylum Research), with NSG30/Au (NT-MDT) probes in tapping mode in air. These golden silicon probes have a nominal resonance frequency and spring constant of 240-440 kHz and 22-100 N/m, respectively. The tip is a pyramid with 14-16 μm length, ~20 nm apex diameter. The images are processed with the AFM company software Version-13, based on IgorPro 6.22 (Wavemetrics). $MoS_2$ is deposited onto FTO as described in the previous section (Fabrication techniques). Images of $MoS_2$ flakes deposited onto mica sheets (EMS) (V-1 quality) are also acquired in order to extrapolate the average thickness of single flakes. $MoS_2$ flakes are deposited by drop casting from 0.1 mg mL$^{-1}$ $MoS_2$ dispersion in IPA. Statistical analysis of the height profiles is carried out by means of software Origin 8.1 (OriginLab).

Work function data are obtained by using a commercial kelvin probe system (KPSP020, KP Technologies Inc.). Samples are measured in air and at room temperature and both a clean gold surface (WF = 4.8 eV) and a graphite sample (HOPG, highly ordered pyrolytic graphite, WF = 4.6 eV) are used as independent references for the probe potential.

### 4.4. Photoelectrochemical measurements

Photoelectrochemical measurements are carried out at room temperature in a flat-bottom fused silica cell under a three-electrode configuration using CompactStat potentiostat/galvanostat station (Ivium), controlled via Ivium's own IviumSoft. A Pt wire is used as the counter-electrode and sat. KCl Ag/AgCl is used as the reference electrode. Measurements are performed in 50 mL aqueous solution of 0.5 M H2SO4 (99.999% purity, Sigma Aldrich) at pH 1. Oxygen is purged from electrolyte solutions by flowing nitrogen gas throughout the liquid volume using a porous frit at least 30 minutes before starting measurements. A constant, slight nitrogen flow is maintained afterwards for the whole duration of experiments, to avoid re-dissolution of molecular oxygen in the electrolyte. Potential differences between the working electrode and the reference electrode are reported with respect to the RHE scale using the Nernst equation. A 300 W Xenon light source LS0306 (Lot Quantum Design), equipped with AM 1.5 G filters, is used to simulate solar illumination (1 sun) at the glass substrate side of the samples inside the test cell. Linear Sweep Voltammetry is used to evaluate the response of devices in the dark and under 1 sun illumination. Voltage is swept starting from the $V_{oc}$ of the photocathodes to a negative potential of -0.3 V *vs.* RHE at a scan rate of 10 mV s$^{-1}$. Stability test is performed by recording in time the $J_{0V\ vs\ RHE}$ under continuous illumination (1 Sun) at 0 V *vs.* RHE. Cyclic voltammetry (CV) is carried out on $MoS_2$ films deposited on glassy carbon substrates (Sigma Aldrich) (previously cleaned with IPA) by consequent drop casting steps from 0.4 mg mL$^{-1}$ $MoS_2$ dispersion in IPA (until reaching a loading of $MoS_2$ of 0.5 mg cm$^{-1}$). The voltage scan rate is 200 mV s$^{-1}$.

## Acknowledgements

We acknowledge M. Prato for useful discussion. This project has received funding from the European Union's Horizon 2020 research and innovation program under grant agreement No. 696656—GrapheneCore1, and from the FP7 FET Collaborative Project PHOCS, under grant agreement n. 309223.

# Supporting information

## S.1. TEM analysis of the MoS$_2$ flakes

In Fig. 1a of the main text of the manuscript is reported a representative transmission electron microscopy (TEM) image of MoS$_2$ flakes, showing irregular shaped sheets with lateral size in the 30-800 nm range (mean value ~275 nm, *i.e*. mean area approx. 0.075 um$^2$) (see statistical analysis of the TEM images reported in Fig. 1B). Fig. S1 report additional TEM images to better illustrate the morphology of MoS$_2$ flakes evidenced by Fig. 1a.

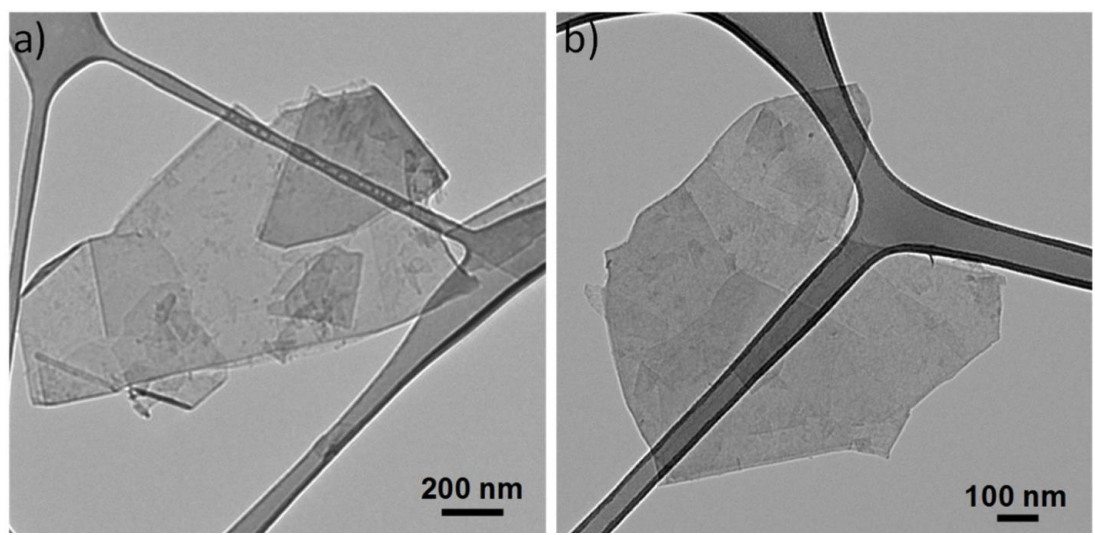

**Fig. S1** a) and b) TEM image of the MoS$_2$ flakes deposited onto carbon coated Cu TEM grids.

## S.2. AFM analysis of MoS$_2$ flakes

In Fig. 1c of the main text of the manuscript we show a representative AFM image of MoS$_2$ flakes with thickness of ~1.2 nm. Here, Fig. S2a shows an AFM image of MoS$_2$ flakes with thickness beetween 3 and 6 nm, while Fig. S2b shows an AFM image of a MoS$_2$ flake with a thickness up to 5.8 nm. Representative flakes' height profiles are shown in Fig. S2 by red lines.

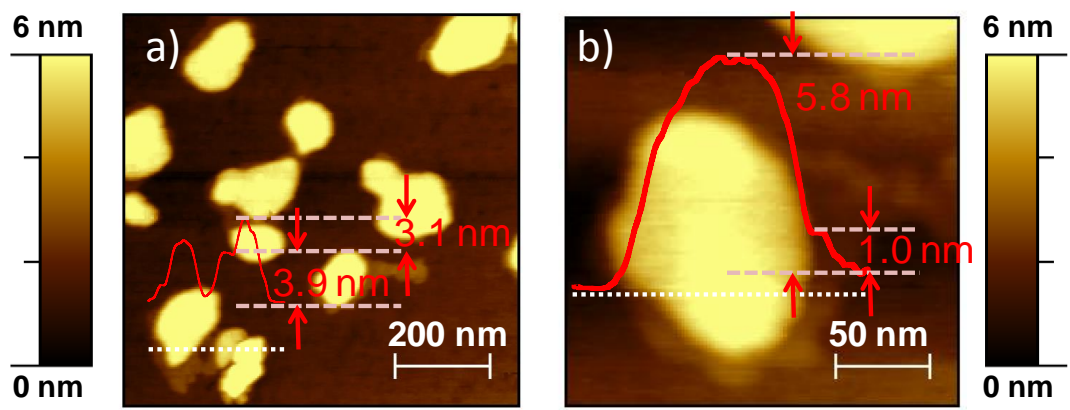

**Fig. S2** a) AFM images of MoS$_2$ flakes deposited onto a V1-quality mica substrate, showing a multilayer structure (up to ~6 nm in thickness). b) AFM image on a zoomed part of image in panel a, showing an approx. 7-layers MoS$_2$ flake. The height profiles of representative flakes are also shown (red lines).

## S.3. Statistical Raman analysis of the MoS2 flakes

Fig. S3 reports the statistical Raman analysis of the MoS$_2$ flakes for the position, the full width at half maximum (FWHM) and the difference of the positions of the two main vibrational modes of MoS$_2$, i.e. $E_{2g}^1$ and $A_{1g}$, while the Raman features of the bulk MoS$_2$ are summarized in Table S1. The $E_{2g}^1(\Gamma)$ mode of the MoS$_2$ flakes (located at ~380 cm$^{-1}$) exhibits softening with respect to the one of the bulk counterpart (located at 378 cm$^{-1}$), while no difference of the peak positon of the $A_{1g}(\Gamma)$ modes (~403 cm$^{-1}$) is observed. Thus, the frequency difference between $A_{1g}(\Gamma)$ and $E_{2g}^1(\Gamma)$ (24 cm$^{-1}$) of the MoS$_2$ decreases of ~2 cm$^{-1}$ with respect to the bulk MoS$_2$ (26 cm$^{-1}$), see also Fig. 2b in the main text. The full width at half maximum (FWHM) of the $E_{2g}^1(\Gamma)$ and $A_{1g}(\Gamma)$ for the exfoliated MoS$_2$ increase of ~2 cm$^{-1}$ and ~1 cm$^{-1}$, respectively, if compared with the corresponding modes of bulk MoS$_2$. As discussed in the main text of the manuscript, these results confirm the different topological structure of the MoS$_2$ flakes with respect to the bulk MoS$_2$.

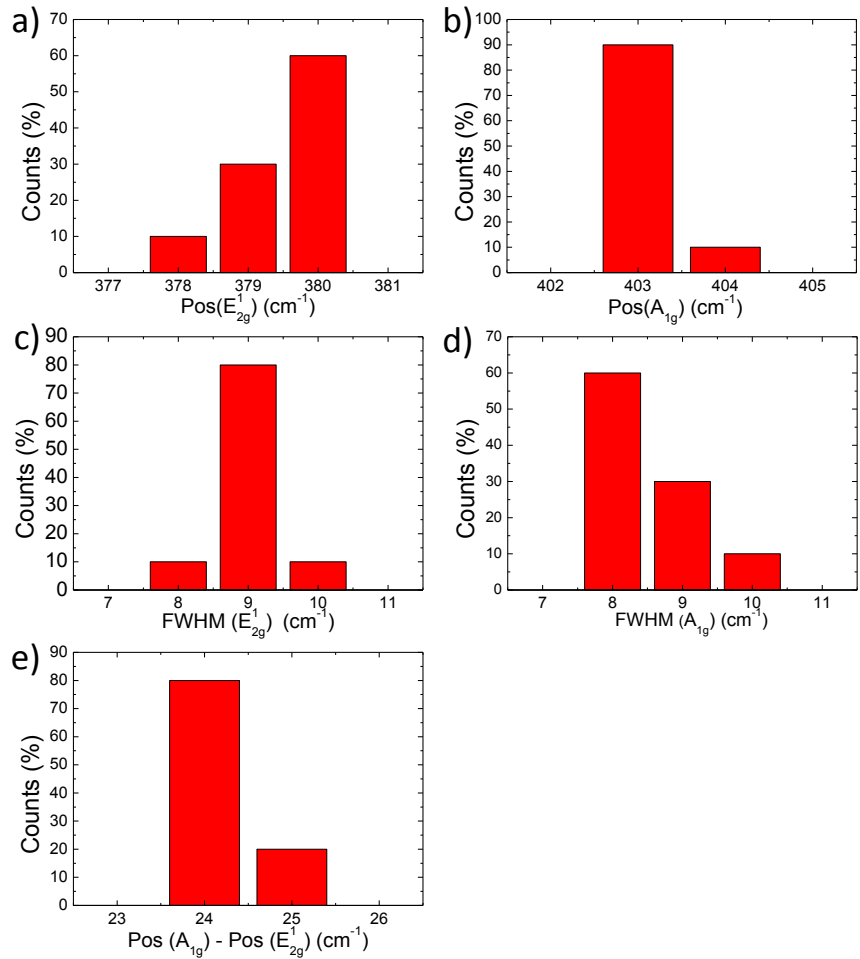

**Fig. S3** Statistical Raman analysis of the MoS$_2$ flakes for a) the position of $E^1_{2g}$ mode, b) the position of $A_{1g}$ mode, c) the FWHM of the $E^1_{2g}$ mode, d) FWHM of $A_{1g}$ mode and e) the relative distance of the difference of the positions of the $A_{1g}$ and $E^1_{2g}$ modes (calculated on 30 different measurements). MoS$_2$ is deposited onto a Si wafer with 300 nm thermally grown SiO$_2$.

**Table S1** Raman characteristics of bulk MoS$_2$.

| POS($E^1_{2g}$) | FWHM($E^1_{2g}$) | POS($A_{1g}$) | FWHM($A_{1g}$) | POS($A_{1g}$)-POS($E^1_{2g}$) |
|---|---|---|---|---|
| 378 | 7 | 403 | 7 | 26 |

### S.4. AFM analysis of FTO/p-MoS$_2$ (10 mM)

Fig. S4 reports the AFM images for the FTO/p-MoS$_2$ (10 mM), which shows no differences in surface morphology with respect to the undoped case (FTO/MoS$_2$), whose image is shown in Fig. 3b of the main text. As shown in Table 2, the Ra of FTO/p-MoS$_2$ (10 mM) (Ra = 11.9 nm) has a similar value to that of FTO/MoS$_2$ (Ra = 11.6 nm). These data suggest that the p-doping treatment up to 10 mM does not affect the surface morphology of the deposited MoS$_2$ films. These results are in agreement with the SEM analysis of Fig. 4b-c of the main text of the manuscript.

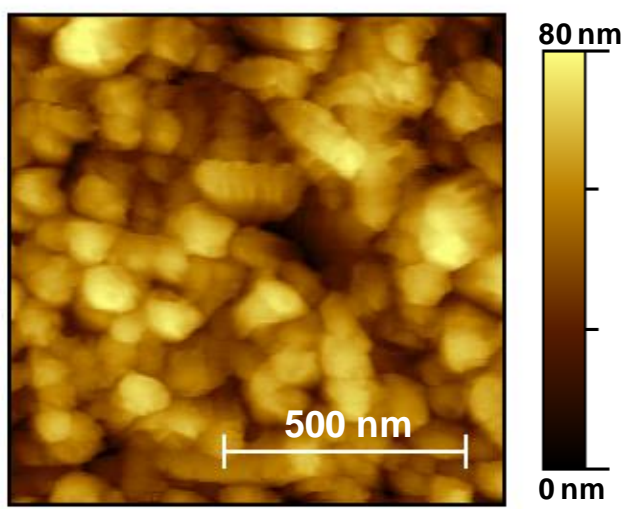

**Fig. S4** AFM images of FTO/p-MoS$_2$ (10 mM). The MoS$_2$ films are deposited from a 0.1 mg mL$^{-1}$ MoS$_2$ dispersion in IPA. The calculated roughness average (Ra) is 11.9 nm (see Table 2 in the main text of the manuscript).

## S.5. Photoelectrochemical characterization of MoS$_2$-based photocathodes fabricated using different concentration of MoS$_2$ dispersion in IPA

Fig. S5 shows the photoelectrochemical responses of the MoS$_2$-based photocathodes fabricated using different concentration (0.05, 0.1, 0.4 mg mL$^{-1}$, dark cyan, orange and green lines) of MoS$_2$ dispersion in isopropanol (IPA) during the deposition of the MoS$_2$ films onto FTO. These results reveal that high photocurrent values are obtained for photocathodes with MoS$_2$ films obtained from 0.1 mg mL$^{-1}$ concentration.

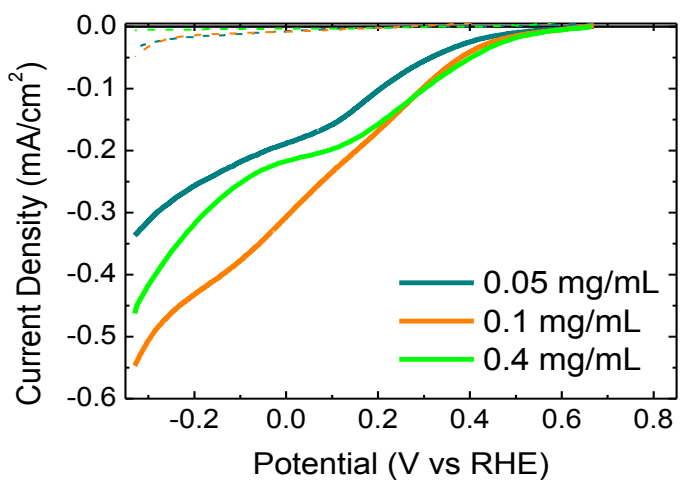

**Fig. S5** linear sweep voltammetry measurements in 0.5 M H$_2$SO$_4$ solution at pH 1, under dark (dashed lines) and AM1.5 light illumination (100 mW cm$^{-2}$) (solid lines) for MoS$_2$-based photocathodes fabricated using different concentration (0.05, 0.1, 0.4 mg mL$^{-1}$, dark cyan, orange and green lines) of MoS$_2$ dispersion in IPA.

## S.6. Photoelectrochemical characterization of p-MoS$_2$ (10 mM)-based photocathodes fabricated using different concentration of MoS$_2$ dispersion in IPA

Fig. S6 shows the photoelectrochemical responses of the p-MoS$_2$ (10 mM)-based photocathodes fabricated using different concentration (0.05, 0.1, 0.4 mg mL$^{-1}$, cyan, red and yellow lines) of MoS$_2$ dispersion in IPA for the deposition of the MoS$_2$ films onto FTO. As for the undoped MoS$_2$-based photocathodes (see Fig. S5), these results reveal that high photocurrent values are obtained for photocathodes with MoS$_2$ films obtained from 0.1 mg mL$^{-1}$ concentration, whose analysis is reported in the main text of the manuscript.

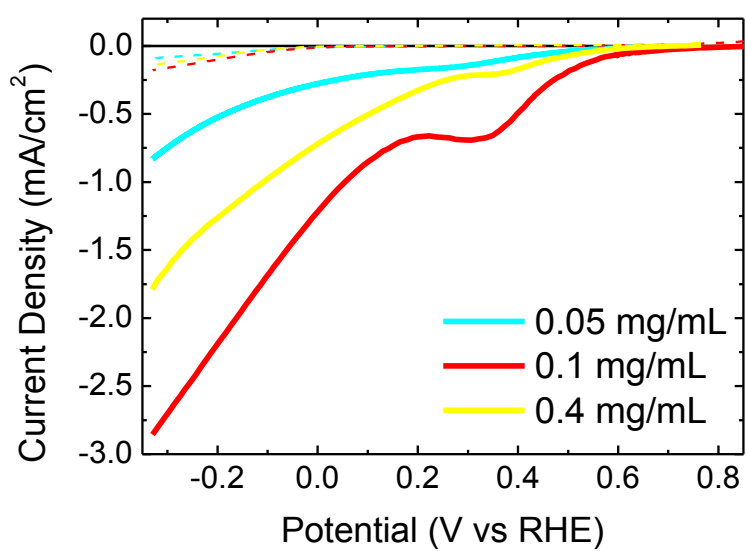

**Fig. S6** Linear sweep voltammetry measured in 0.5 M H$_2$SO$_4$ solution at pH 1, under dark (dashed lines) and AM1.5 light illumination (100 mW cm$^{-2}$) (solid lines) for p-MoS$_2$ (10 mM)-based photocathodes fabricated using different concentration (0.05, 0.1, 0.4 mg mL$^{-1}$, cyan, red and yellow lines) of MoS$_2$ dispersion in IPA.

## S.7. Linear sweep voltammetry (LSV) measurements of FTO/MoS$_3$

Fig. S7 reports two consecutive LSVs of FTO/MoS$_3$ measured in 0.5 M H$_2$SO$_4$ solution at pH 1, which is the same testing conditions of the LSVs reported in Fig. 6 for the photocathodes studied in this work. It is evident that the first LSV exhibits a broad reduction peak (marked by dashed blue oval) at positive potentials with respect to the HER (located at ~-0.2 V *vs.* RHE). The peak is reduced during the second LSV, which also shows a shift of the HER activity towards more positive potentials, as indicated by the blue arrow (see also the inset panel where LSV scan are showed over a negatively extended range of potential). These results confirm that during the first LSV scans the MoS$_3$ is reduced towards more catalytic MoS$_{2+x}$ species, as also studied by other works (see ref. 68(b) of the main manuscript).

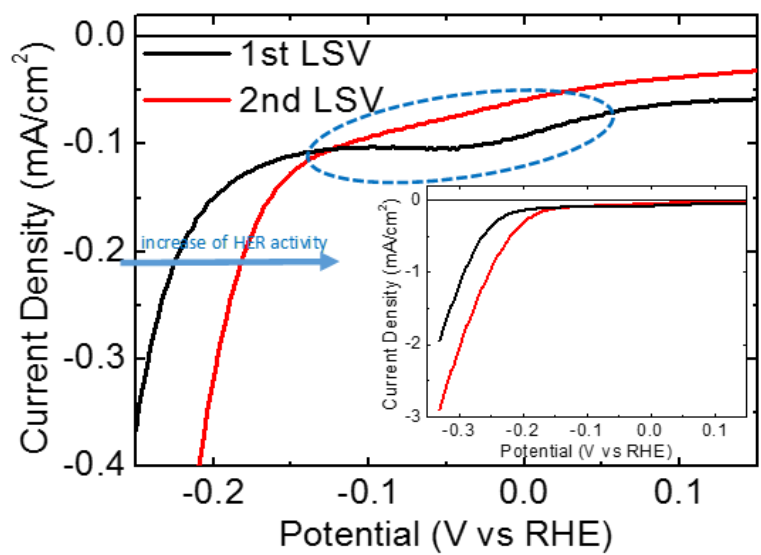

**Fig. S7** Consecutive LSVs of FTO/MoS$_3$ measured in 0.5 M H$_2$SO$_4$ solution at pH 1. The broad reduction peak at potentials more positive than the HER (located at ~-0.2 V *vs.* RHE) is marked dashed blue oval and it is more pronounced during the first LSV. The peak is reduced during the second LSV, which also shows a shift of the HER activity (as evidenced by the blue arrow). The inset panel show the same LSVs of the main panel, extended over larger J and V scales.

### S.8. Cyclic Voltammetry measurements on MoS$_2$ films

Fig. S8 reports Moreover, cyclic voltammetry (CV) measurements of MoS$_2$ films deposited on glassy carbon (GC) substrates by consequent drop casting steps from 0.4 mg mL$^{-1}$ MoS$_2$ dispersion in IPA until reaching a loading of MoS$_2$ of 0.5 mg cm$^{-1}$. The figure also reports the CV of GC as reference background, revealing its electrochemical inertness. These data show the absence of irreversible redox reaction involving MoS$_2$ films under the HER-working condition of the photocathodes reported in Fig. 6 of the main text of the manuscript, in agreement with the electrochemical stability of the MoS$_2$ expressed in several energy conversion and storage devices, including fuel and water splitting cells, batteries and supercapacitors (see ref. 70-73 in the main text of the manuscript). In the inset to Fig. S8 the HER-electrocatalytic activity of the MoS$_2$ is also reported by LSV measurement over more negative voltages (up to -0.8 V vs RHE), showing onset over-potential of ~200 mV with respect to the RHE potential, in accordance with previous studies on layered MoS$_2$ materials (see ref. 71 in the main test of the manuscript). It is worth to note that this value is similar to that obtained for the MoS$_3$ (180 mV) (see Fig. 6 in the main text) used as EC in the photocathodes here studied, suggesting its possible exploitation in the photocathodes' structures as well as EC.

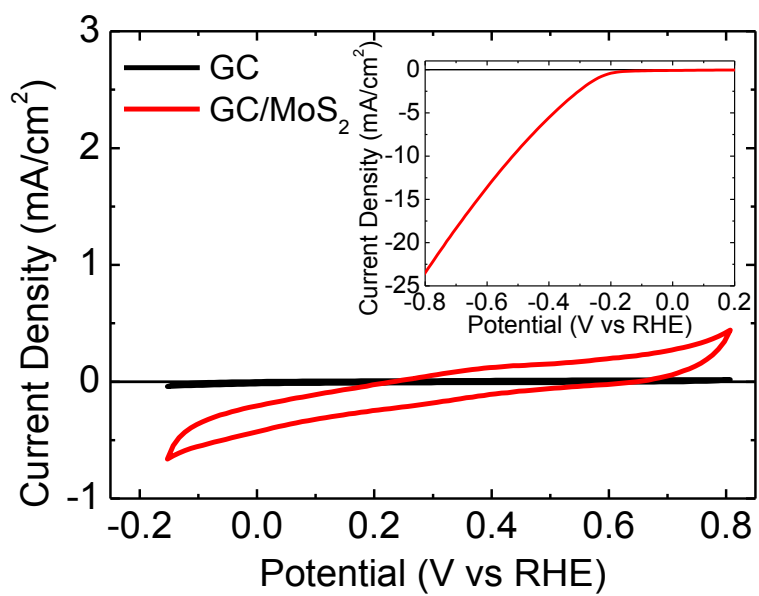

**Fig. S8** CVs measured for GC and GC/MoS$_2$ samples in in 0.5 M H$_2$SO$_4$ solution at pH 1. Voltage scan rate: 500 mV s$^{-1}$. MoS$_2$ films are deposited on glassy carbon (GC) substrates by consequent drop casting steps from 0.4 mg mL$^{-1}$ MoS$_2$ dispersion in IPA until reaching a loading of MoS$_2$ of 0.5 mg cm$^{-1}$. The inset panel show the HER-electrocatalytic activity of MoS$_2$ flakes, as measured by LSV on GC/MoS$_2$ sample.